\documentclass[12pt]{article}

\usepackage{amsmath}
\usepackage{amssymb}
\usepackage{amsfonts}
\usepackage{graphics}

\title{
Generalized semiconfined harmonic oscillator model with a position-dependent effective mass}
\author{C. Quesne\thanks{e-mail: Christiane.Quesne@ulb.be}\\
{\small\sl Physique Nucl\'eaire Th\'eorique et Physique Math\'ematique,  Universit\'e Libre de Bruxelles,} \\ 
{\small\sl Campus de la Plaine CP229, Boulevard~du Triomphe, B-1050 Brussels, Belgium}}
\date{ }
\begin{document}
\baselineskip=22pt plus 1pt minus 1pt
\maketitle

\begin{abstract} 
By using a point canonical transformation starting from the constant-mass Schr\"odinger equation for the isotonic potential, it is shown that a semiconfined harmonic oscillator model with a position-dependent mass in the BenDaniel-Duke setting and the same spectrum as the standard harmonic oscillator can be easily constructed and extended to a semiconfined shifted harmonic oscillator, which could result from the presence of a uniform gravitational field. A further generalization is proposed by considering a $m$-dependent position-dependent mass for $0<m<2$ and deriving the associated semiconfined potential. This results in a family of position-dependent mass and potential pairs, to which the original pair belongs as it corresponds to $m=1$. Finally, the potential that would result from a general von Roos kinetic energy operator is presented and the examples of the Zhu-Kroemer and Mustafa-Mazharimousavi settings are briefly discussed.
\end{abstract}


%
%
\newpage

\section{Introduction}

There is much interest in the Schr\"odinger equation wherein the constant mass is replaced by a position-dependent mass (PDM), because the latter has many applications in problems occurring in several fields of physics \cite{bastard, weisbuch, serra, harrison, barranco, geller, arias, puente, ring, bonatsos, willatzen, chamel}. As it has been shown \cite{cq04} that the PDM Schr\"odinger equation is equivalent to two other unconventional Schr\"odinger equations, namely the Schr\"odinger equation resulting from the use of deformed commutation relations \cite{kempf, hinrichsen, witten}, as well as that in curved space \cite{schro, kalnins96, kalnins97}, this has reinforced the interest in its study.\par
%
%
As a consequence, much attention has been devoted to finding exact solutions of PDM Schr\"odinger equations because they may provide a conceptual understanding of some physical phenomena, as well as a testing ground for some approximation schemes. The generation of PDM and potential pairs leading to such exact solutions has been achieved by various methods (see, e.g., \cite{cq06} and references quoted therein). One of the most powerful techniques for such a purpose consists in applying a point canonical transformation (PCT) to an exactly solvable constant-mass Schr\"odinger equation \cite{bagchi, cq09}. Recently, such an approach has proved its efficiency again by providing a straightforward generalization \cite{cq21} of a harmonic oscillator model wherein both the mass and the angular frequency are dependent on the position \cite{jafarov20a}.\par
%
%
The purpose of the present paper is to re-examine a new model of semiconfined harmonic oscillator with a mass that varies with position, which has the striking property of having the same spectrum as the standard harmonic oscillator model \cite{jafarov21a}. By using the PCT method, we plan to prove that one can find a family of PDM and semiconfined potential pairs corresponding to such a spectrum and to which the original PDM and semiconfined harmonic oscillator pair belongs.\par
%
%
The paper is organized as follows. In Sec.~2, the model of \cite{jafarov21a} is reviewed and shown to be derivable by applying the PCT technique to the constant-mass isotonic oscillator model \cite{weissman, zhu}. In Sec.~3, an extension of the model is proposed by starting from a more general PDM and determining the associated semiconfined potential. Finally, Sec.~4 contains some comments.\par
%
%
\section{Semiconfined harmonic oscillator model and its derivation by the PCT technique}

Jafarov and Van der Jeugt recently determined the exact solution of a PDM semiconfined harmonic oscillator model, characterized by the Schr\"odinger equation \cite{jafarov21a}
\begin{equation}
  \left(- \frac{d}{dx} \frac{1}{M(x)} \frac{d}{dx} + V_{\rm eff}(x)\right) \psi_n(x) = E_n \psi_n(x),
  \label{eq:SE-PDM}
\end{equation}
where the kinetic energy operator has the BenDaniel-Duke form \cite{bendaniel} and the potential has the harmonic oscillator form
\begin{equation}
  V_{\rm eff}(x) = \frac{1}{4} M(x) \omega^2 x^2,  \label{eq:V-ho}
\end{equation}
except that the mass
\begin{equation}
  M(x) = \begin{cases}
     \left(1+\frac{x}{a}\right)^{-1} & \text{if $-a<x<+\infty$}, \\[0.2cm]
     +\infty  & \text{if $x\le -a$},   
  \end{cases} \label{eq:M}
\end{equation}
with $a>0$, depends on the position in such a way that $V_{\rm eff}(-a) = +\infty$ and $\lim_{x\to +\infty} V_{\rm eff}(x) = + \infty$.\footnote{Note that we have adopted here units wherein $\hbar = 2m_0$ in the original paper.}\par
%
%
By directly solving the differential equation (\ref{eq:SE-PDM}), they found that the spectrum of this semiconfined model is that of the standard harmonic oscillator,
\begin{equation}
  E_n = \omega \left(n+\frac{1}{2}\right), \qquad n=0, 1, 2, \ldots,  \label{eq:E}
\end{equation}
with corresponding wavefunctions
\begin{equation}
  \psi_n(x) = C_n \left(1+\frac{x}{a}\right)^{\frac{1}{2}\omega a^2} e^{-\frac{1}{2}\omega a(x+a)}
  L_n^{(\omega a^2)}\left(\omega a^2 \left(1+\frac{x}{a}\right)\right), \qquad -a<x<+\infty,
\end{equation}
expressed in terms of Laguerre polynomials $L_n^{(\alpha)}(z)$ and vanishing at $x=-a$ and $x\to+\infty$, as it should be. Here $C_n$ is a normalization coefficient given by
\begin{equation}
  C_n = (\omega a^2)^{\frac{1}{2}(\omega a^2 +1)} \sqrt{\frac{n!}{a \Gamma(\omega a^2 + n +1)}}.
\end{equation}
\par
%
%
These results may be alternatively derived by applying a PCT to the constant-mass Schr\"odinger equation for the isotonic oscillator \cite{weissman, zhu}
\begin{equation}
  \left(- \frac{d^2}{du^2} + U(u)\right) \phi_n(u) = \epsilon_n \phi_n(u),  \label{eq:SE}
\end{equation}
where
\begin{equation}
  U(u) = \frac{1}{4} \bar{\omega}^2 u^2 + \frac{g}{u^2}, \qquad g>0, \qquad 0<u<+\infty,  \label{eq:U} 
\end{equation}
\begin{equation}
  \epsilon_n = \bar{\omega} (2n + \alpha + 1), \qquad \alpha = \frac{1}{2} \sqrt{1 + 4g},   \label{eq:alpha} 
\end{equation}
and 
\begin{equation}
  \phi_n(u) \propto u^{\alpha+\frac{1}{2}} e^{-\frac{1}{4} \bar{\omega} u^2} L_n^{(\alpha)}\left(\frac{1}{2}
  \bar{\omega} u^2\right).  \label{eq:phi}
\end{equation}
\par
%
%
A PCT transforming an equation such as (\ref{eq:SE}) into a PDM equation of type (\ref{eq:SE-PDM}) \cite{bagchi, cq09} consists in making a change of variable
\begin{equation}
  u(x) = \bar{a} v(x) + \bar{b}, \qquad v(x) = \int^x \sqrt{M(x')}\, dx',  \label{eq:u}
\end{equation}
and a change of function
\begin{equation}
  \phi_n(u(x)) \propto [M(x)]^{-1/4} \psi_n(x).  \label{eq:phi-psi}
\end{equation}
The potential $V_{\rm eff}(x)$ and the energy eigenvalues $E_n$ of the PDM Schr\"odinger equation are then given in terms of the potential and the energy eigenvalues of the constant-mass one by
\begin{equation}
  V_{\rm eff}(x) = \bar{a}^2 U(u(x)) + \frac{M^{\prime\prime}}{4M^2} - \frac{7M^{\prime2}}{16M^3} + \bar{c},
  \label{eq:V-U}
\end{equation}
and
\begin{equation}
  E_n = \bar{a}^2 \epsilon_n + \bar{c},
\end{equation}
where a prime denotes derivation with respect to $x$ and $\bar{a}$, $\bar{b}$, $\bar{c}$ are three real constants.\par
%
%
In the present case, from (\ref{eq:M}) and (\ref{eq:u}), we directly obtain
\begin{equation}
  v(x) = 2a \sqrt{1+\frac{x}{a}}. \label{eq:v}
\end{equation}
and
\begin{equation}
  \frac{M^{\prime\prime}}{4M^2} - \frac{7M^{\prime2}}{16M^3} = \frac{1}{16a^2} \left(1+\frac{x}{a}
  \right)^{-1} \label{eq:MM}
\end{equation}
for $-a<x<+\infty$. With the choice $\bar{a} = \sqrt{\frac{\omega}{2\bar{\omega}}}$, $\bar{b}=0$, we get for the change of variable (\ref{eq:u})
\begin{equation}
  u(x) = a \sqrt{\frac{2\omega}{\bar{\omega}}} \sqrt{1 + \frac{x}{a}}
\end{equation}
and the change of function (\ref{eq:phi-psi}), together with (\ref{eq:phi}), leads to
\begin{equation}
  \psi_n(x) = C_n \left(1+ \frac{x}{a}\right)^{\alpha/2} e^{-\frac{1}{2}\omega a(x+a)}
  L_n^{(\alpha)}\left(\omega a^2 \left(1+\frac{x}{a}\right)\right),
\end{equation}
where $C_n$ turns out to be
\begin{equation}
  C_n = (\omega a^2)^{\frac{1}{2}(\alpha+1)} \sqrt{\frac{n!}{a \Gamma(\alpha+n+1)}}.
\end{equation}
Furthermore, on assuming $\bar{c} = - \frac{\omega}{2}\alpha$, the transformed potential (\ref{eq:V-U}) becomes
\begin{equation}
  V_{\rm eff}(x) = \frac{a\omega^2}{4(x+a)} \left(x+a-\frac{\alpha}{a\omega}\right)^2, \label{eq:V}
\end{equation}
with corresponding eigenvalues $E_n$ given by (\ref{eq:E}).\par
%
%
If we compare these results with those of \cite{jafarov21a}, we notice that we have obtained the same energy spectrum (\ref{eq:E}), but with generalized potential and wavefunctions, since the latter depend on an extra parameter $\alpha$ absent in \cite{jafarov21a}. By taking $\alpha = a^2 \omega$, the original results are retrieved, but for other values of $\alpha$, the potential (\ref{eq:V}) describes a semiconfined shifted harmonic oscillator. Note that such a potential might be interpreted as a semiconfined harmonic oscillator in a uniform gravitational field as was done for a shifted harmonic oscillator with another type of PDM \cite{jafarov20b, jafarov21b}.\par
%
%
\section{Family of generalized semiconfined oscillator models}

A further generalization of the model of \cite{jafarov21a} can be obtained by changing the PDM (\ref{eq:M}) into a PDM depending on some parameter $m$ taking values in the interval $0<m<2$,
\begin{equation}
  M(x) = \begin{cases}
       \left(1+\frac{x}{a}\right)^{-m} & \text{if $-a<x<+\infty$}, \\[0.2cm]
     +\infty  & \text{if $x\le -a$},   
  \end{cases} \label{eq:M-m}
\end{equation}
and determining the associated potential $V_{\rm eff}(x)$ with the assumption that the starting  constant-mass Schr\"odinger equation remains as given in (\ref{eq:SE}) and (\ref{eq:U}). The results of Sec.~2 will then correspond to the $m=1$ special case.\par
%
%
Equations (\ref{eq:v}) and (\ref{eq:MM}) are now replaced by
\begin{equation}
  v(x) = \frac{2a}{2-m} \left(1 + \frac{x}{a}\right)^{1-\frac{m}{2}}
\end{equation}
and
\begin{equation}
  \frac{M^{\prime\prime}}{4M^2} - \frac{7M^{\prime2}}{16M^3} = - \frac{1}{16a^2} m(3m-4) \left(1+
  \frac{x}{a}\right)^{m-2},
\end{equation}
respectively. On keeping the same values for $\bar{a}$, $\bar{b}$, and $\bar{c}$ as in Sec.~2, we get a new change of variable
\begin{equation}
  u(x) = \frac{a}{2-m} \sqrt{\frac{2\omega}{\bar{\omega}}} \left(1+\frac{x}{a}\right)^{1-\frac{m}{2}},
\end{equation}
but the resulting energy eigenvalues remain given by (\ref{eq:E}). From (\ref{eq:V-U}), however, the resulting potential turns out to be $m$-dependent and given by
\begin{align}
  V_{\rm eff}(x) &= \frac{a^m \omega^2}{4(2-m)^2} (x+a)^{2-m} + \frac{[(m-2)\alpha-(m-1)][(m-2)\alpha
  +m-1]}{4a^m(x+a)^{2-m}} \nonumber \\
  &\quad {} - \frac{1}{2}\omega \alpha.  \label{eq:V-gen}
\end{align}
This is also the case for the wavefunctions, which become
\begin{align}
  \psi_n(x) &= C_n \left(1+\frac{x}{a}\right)^{-\frac{m}{2}(\alpha+1)+\alpha+\frac{1}{2}}
      e^{- \frac{1}{2(2-m)^2} \omega a^2 \left(1+\frac{x}{a}\right)^{2-m}} \nonumber \\
  &\quad {} \times L_n^{(\alpha)}\left(\frac{\omega a^2}{(2-m)^2} \left(1+\frac{x}{a}\right)^{2-m}\right),
      \label{eq:psi-gen}
\end{align}
where
\begin{equation}
  C_n = \left(\frac{\omega a^2}{(2-m)^2}\right)^{\frac{1}{2}(\alpha+1)} \sqrt{\frac{(2-m) n!}{a
  \Gamma(\alpha+n+1)}}.
\end{equation}
\par
%
%
The new $m$-dependent potential (\ref{eq:V-gen}) will be a semiconfined potential provided it goes to $+\infty$ for $x\to +\infty$ and $x\to -a$. The former condition is automatically satisfied, but the latter imposes that
\begin{equation}
  \alpha > \frac{m-1}{2-m},
\end{equation}
which implies a restriction for $m$ values such that $\frac{m-1}{2-m} > \frac{1}{2}$, i.e., for those in the interval $\frac{4}{3}<m<2$. In such a case, the wavefunctions (\ref{eq:psi-gen}) vanish for $x\to +\infty$ and $x\to -a$, as it should be. The minimum of the potential occurs for
\begin{equation}
  x_{\rm min} = -a + \left\{\frac{(2-m)^2}{a^m \omega} \sqrt{\alpha^2 - \left(\frac{m-1}{2-m}\right)^2}
  \right\}^{1/(2-m)}  \label{eq:xmin}
\end{equation}
and is given by
\begin{equation}
  (V_{\rm eff})_{\rm min} = \frac{1}{2}\omega \left\{\sqrt{\alpha^2 - \left(\frac{m-1}{2-m}\right)^2}
  - \alpha\right\}.  \label{eq:Vmin}
\end{equation}
It is therefore slightly negative, except for $m=1$ for which it vanishes.\par
%
%
In Fig.~1, we show the dependence of the semiconfined potential (\ref{eq:V-gen}) on $m$. The black line corresponds to the original semiconfined harmonic oscillator (\ref{eq:V-ho}).\par
%
%
\begin{figure}
\begin{center}
\includegraphics{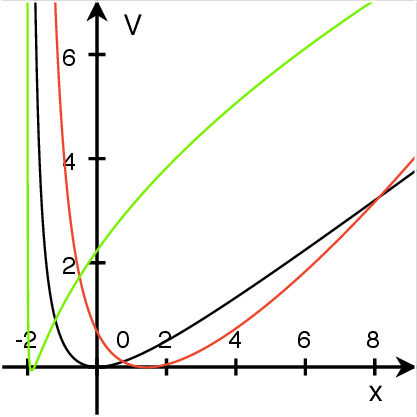}
\caption{Plot of the semiconfined potential (\ref{eq:V-gen}) in terms of $x$ for $m=1$ (black line), $m=\frac{1}{2}$ (red line), and $m=\frac{3}{2}$ (green line). The parameter values are $\omega=1$, $a=2$, and $\alpha=4$.}
\end{center}
\end{figure}
\par
%
%
\section{Comments}

In the present paper, we have first shown that the PCT method applied to the constant-mass Schr\"odinger equation for the isotonic oscillator allows us to easily retrieve the results of \cite{jafarov21a} and to extend them in order to describe a semiconfined shifted harmonic oscillator, which might be interpreted as a semiconfined harmonic oscillator in a uniform gravitational field.\par
%
%
In a second step, we have obtained a further generalization by considering a $m$-dependent PDM for $0<m<2$ and by deriving the corresponding semiconfined potential with the same spectrum as the standard harmonic oscillator. We have therefore constructed a family of PDM and potential pairs, to which the original pair belongs as it corresponds to $m=1$.\par
%
%
In \cite{jafarov21a}, the BenDaniel-Duke ordering \cite{bendaniel} was chosen for the momentum and mass operators. One finds, however, in the literature, several other orderings, which are special cases of the von Roos general two-parameter form of the kinetic energy operator \cite{vonroos}, for which the Schr\"odinger operator writes
\begin{align}
  &\biggl\{- \frac{1}{2} \biggl[M(x)^{\xi} \frac{d}{dx} M(x)^{\eta} \frac{d}{dx} M(x)^{\zeta} + M(x)^{\zeta}
  \frac{d}{dx} M(x)^{\eta} \frac{d}{dx} M(x)^{\xi}\biggr] + V_{\rm vR}(x)\biggr\} \psi_n(x)\nonumber \\
  &\quad{}  = E_n \psi_n(x),  \label{eq:SE-PDM-bis}
\end{align}
where $\xi$, $\eta$, $\zeta$ are some real parameters restricted by the condition $\xi + \eta + \zeta = -1$. In particular, the BenDaniel-Duke ordering corresponds to $\xi = \zeta =0$, $\eta = -1$ and the relation between the potentials in (\ref{eq:SE-PDM}) and (\ref{eq:SE-PDM-bis}) is given by
\begin{equation}
  V_{\rm vR}(x) = V_{\rm eff}(x) - \frac{1}{2}(1+\eta) \frac{M^{\prime\prime}}{M^2} + [\xi(\xi+\eta+1)
  +\eta+1] \frac{M^{\prime2}}{M^3}.
\end{equation}
For the mass chosen in (\ref{eq:M-m}), the latter becomes
\begin{align}
  & V_{\rm vR}(x) \nonumber \\
  &= V_{\rm eff}(x) + \left\{- \frac{1}{2}(1+\eta) m(m+1) + [\xi(\xi+\eta+1)+\eta+1] m^2\right\}
      \frac{(x+a)^{m-2}}{a^m}.  \label{eq:vonroos}
\end{align}
\par
%
%
It is worth noting, in particular, the Zhu-Kroemer \cite{ZK} and Mustafa-Mazharimousavi \cite{MM} orderings, which pass the de Souza Dutra and Almeida test \cite{dutra} as good orderings. The former corresponds to $\xi=\zeta= - \frac{1}{2}$, $\eta=0$, and leads to replacing (\ref{eq:vonroos})  by
\begin{equation}
  V_{\rm ZK}(x) = \frac{a^m \omega^2}{4(2-m)^2} (x+a)^{2-m} + \frac{[(m-2)\alpha-1][(m-2)\alpha+1]}
  {4a^m(x+a)^{2-m}} - \frac{1}{2}\omega\alpha,
\end{equation}
while the latter is associated with $\xi=\zeta=-\frac{1}{4}$, $\eta=-\frac{1}{2}$, and gives rise to
\begin{equation}
  V_{\rm MM}(x) = \frac{a^m \omega^2}{4(2-m)^2} (x+a)^{2-m} + \frac{(m-2)^2(\alpha^2-\frac{1}{4})}
  {4a^m(x+a)^{2-m}} - \frac{1}{2}\omega\alpha.
\end{equation}
These potentials have a behaviour very similar to that of $V_{\rm eff}(x)$, since they are semiconfined for $\alpha$ restricted to $\alpha>1/(2-m)$ or for any value of $\alpha$ ($>1/2$ by definition (\ref{eq:alpha})), respectively. The place of the minimum and its value are given by (\ref{eq:xmin}) and (\ref{eq:Vmin}) provided $\sqrt{\alpha^2 - [(m-1)/(2-m)]^2}$ is replaced by $\sqrt{\alpha^2 - 1/(2-m)^2}$ or $\sqrt{\alpha^2 - 1/4}$.\par 
%
%
\section*{Acknowledgements}

This work was supported by the Fonds de la Recherche Scientifique-FNRS under Grant No.~4.45.10.08.\par
%
%

\end{document}